\documentclass[fleqn,usenatbib]{mnras}

\usepackage{newtxtext,newtxmath}

\usepackage[T1]{fontenc}
\usepackage{ae,aecompl}

\usepackage{siunitx}
\usepackage{dcolumn}

\usepackage{graphicx}	
\usepackage{amsmath}	
\usepackage{amssymb}	
\usepackage{color}

\DeclareMathOperator\erf{erf}

\title[VPF in BOSS]{Probing Galaxy Assembly Bias in BOSS Galaxies
  Using Void Probabilities}

\author[Walsh et al.]{
Kilian Walsh,$^{1}$
Jeremy Tinker$^{1}$\thanks{e-mail: jeremy.tinker@nyu.edu}
\\
$^{1}$Center for Cosmology and Particle Physics, New York University, New York\\
}

\date{Accepted XXX. Received YYY; in original form ZZZ}

\pubyear{2019}

\begin{document}
\label{firstpage}
\pagerange{\pageref{firstpage}--\pageref{lastpage}}
\maketitle

\begin{abstract}
We measure the void probability function (VPF) of galaxies in
the Baryon Oscillation Spectroscopic Survey (BOSS). The VPF provides
complementary information to standard two-point statistics in that it
is sensitive to galaxy bias in the most extreme underdensities in the
cosmic web. Thus the VPF is ideal for testing whether halo occupation
of galaxies depends on large-scale density, an effect known as galaxy
assembly bias. We find that standard HOD model---one parameterized by
halo mass only---fit only to the two-point function, accurately
predicts the VPF. Additionally, for HOD models where density
dependence is explicitly incorporated, the best-fit models fit to the
combination of the correlation function and the VPF have zero density
dependence. Thus galaxy assembly bias is not a strong source of
systematic uncertaintiy when modeling the clustering of massive galaxies.

\end{abstract}




\section{Introduction}

The growth of structure in the dark matter content of the universe
leads to the formation of regions of high relative density over time
and, contrastingly, also regions of significant sparsity in the matter
field. Galaxies act as a biased tracer of the underlying dark matter
density field (see \citealt{GalBiasRev} for a thorough review of large
scale galaxy bias), so these regions of sparsity lead to the existence
of large voids within the galaxy distribution (see the review of
\citealt{VdW11}). Since the discovery of voids that were larger in
extent than typical galaxy clusters \citep{GregoryThompson}, they have
been considered a probe of interest for the study of cosmology and
large scale structure formation (see e.g. \citealt{BetancortRijo09},
\citealt{Einasto11}, \citealt{Lavaux12}, \citealt{Clampitt16},
\citealt{Kitaura16}, and \citealt{Mao17} for recent examples).

Voids also present a useful probe to understand the nature of galaxy
formation. It is known that galaxies found in voids are typically
spiral/irregularly-shaped and found to be HI rich, having low
luminosity and stellar mass, and small size relative to the general
population of galaxies (\citealt{Tavasoli15}, \citealt{Beygu17}, and
\citealt{Pustilnik19}). This suggests that galaxy formation and
evolution may have a dependence on the large scale dark matter
density. Controlling for factors such as galaxy mass, or the halo
abundances found in underdense regions of simulations, there are also
studies done in the context of voids which claim to find no special
relationship between galaxies and their environment \citep{Patiri06,
  Croton08, Tinker08, Tinker09}.

It is well established, theoretically and in high-resolution dark
matter simulations, that the properties of dark matter halos in
$\Lambda CDM$ cosmologies -- such as halo mass \citep{PressSchechter,
  MoWhite} and formation time \citep{ShethTormen, Wechsler06,
  Zentner07, Dalal08} -- are also strongly dependent on their
large-scale dark matter over density. As such, the observed
correlations of galaxy properties with environment may be explained
with a model that connects galaxies with halos.

One of the chief models of the galaxy-halo connection is the Halo
Occupation Distribution (HOD; see the review of the galaxy-connection
by \cite{wechsler_tinker:18}) The HOD parameterizes the bias between
galaxies and dark matter by the statistical relationship between dark
matter halos and the number of galaxies within them (e.g.,
\citealt{roman_etal:01, berlind_weinberg:02, cooray_sheth:02}).  The
HOD has been remarkably successful in explaining a host of
observational phenomena (e.g., \citealt{Zehavi}, \citealt{Tinker12},
\citealt{Reid14}, and \citealt{Coupon15}).  In its most basic form,
the HOD parameterizes the occupation of a halo by galaxies based only
on the mass of the halo. However, recent studies have suggested that
there may be "assembly bias" in galaxy occupation, whereby the
occupation of dark matter halos may depend on properties other than
mass (\citealt{reddick_etal:13, zentner_etal:16,
  lehmann_etal:17}). Physically, this would manifest as a correlation
between galaxy formation efficiency and that large-scale environment
of a galaxy at fixed halo mass. \cite{tinker_etal:06_voids} proposed
using void statistics, in combination with two-point statistics, to
determine if such a correlation exists. The two-point correlation
function is sensitive to halos in mean and high-density environments,
where most pairs are found, while voids are defined by halos in the
most extreme low-density regions of the universe. If galaxy formation
efficiency changes from high to low densities, one could not
simultaneously fit both statistics using an HOD model that
parameterizes occupation solely on halo mass. \cite{Zentner14} found
similar results using abundance matching models of galaxy bias to
construct color-defined galaxy samples. Using this technique,
\cite{tinker_etal:08_voids} found that the galaxies in the Sloan
Digital Sky Survey (\citealt{sdss}) could be fit with this mass-only
approach. Any correlation of the HOD with large-scale environment
could only be minimal, or else the models could no longer fit both
statistics.

In this work, we extend the investigation of the HOD for a far larger
sample of galaxies, at higher redshifts, by performing this test using
data from the Baryon Oscillation Spectroscopic Survey (BOSS;
\citealt{Boss}). Unlike the SDSS, the BOSS survey is
designed for cosmological inference. Attempts to use BOSS clustering
at non-linear scales require a full model for galaxy bias at those
scales (\citealt{Reid14, rodriguez_torres_etal:16, aemulus3}). If
galaxy assembly bias exists and is not included in the bias model, the
cosmological constraints will themselves be biased
(\citealt{mccarthy_etal:18}). The BOSS galaxy sample also differs from
the low-redshift SDSS samples in that the target selection includes
complicated cuts in galaxy color-color space, and does not yield a
strict volume-limited sample (\citealt{reid_etal:16}).  To perform this
test, we make comparisons with mock measurements made in simulated
galaxy populations derived using the HOD with the high resolution
simulation from the MultiDark collaboration \citep{Riebe13}, MDPL.

The remaining sections of the chapter are organized as follows:
Section~\ref{sec:vpf_dat} describes the BOSS data in detail and the
procedure of making measurements of the 2-point correlation function
and void probability function; Section~\ref{sec:vpf_mod} describes the
HOD model for galaxy occupation of halos, with and without environment
dependence, and how the mock measurements of the correlation function
and VPF are made; Section~\ref{sec:vpf_res} presents our findings for
the HOD model when applied to the BOSS data; and in the final section,
we offer our concluding remarks.

\section{Data \& Measurements}
\label{sec:vpf_dat}

\subsection{BOSS Data}

BOSS \citep{BOSS} is one of the spectroscopic surveys of SDSS-III
\citep{SDSS3}, comprising 1.5 million galaxies and intended to measure
the Baryon Acoustic Oscillation (BAO) feature (see
\citealt{WeinbergReview} for a review). The full survey spans 10,000
square degrees and probes up to redshift $z = 0.8$. Target selection
for the CMASS sample used here is described in
\cite{reid_etal:16}. All results here use Data Release 11 of the SDSS,
the penultimate data release of BOSS, which comprises roughly 85\% of
the full footprint. Additionally, we restrict our analyses to the
North Galactic Cap region (NGC), which is the bulk of the BOSS
footprint, yielding a total sample covering roughly 6,000 deg$^2$.

\subsection{Data Preparation}
\label{sec:dataprep}

\begin{figure}
    \begin{center}
    \includegraphics[width=\columnwidth]{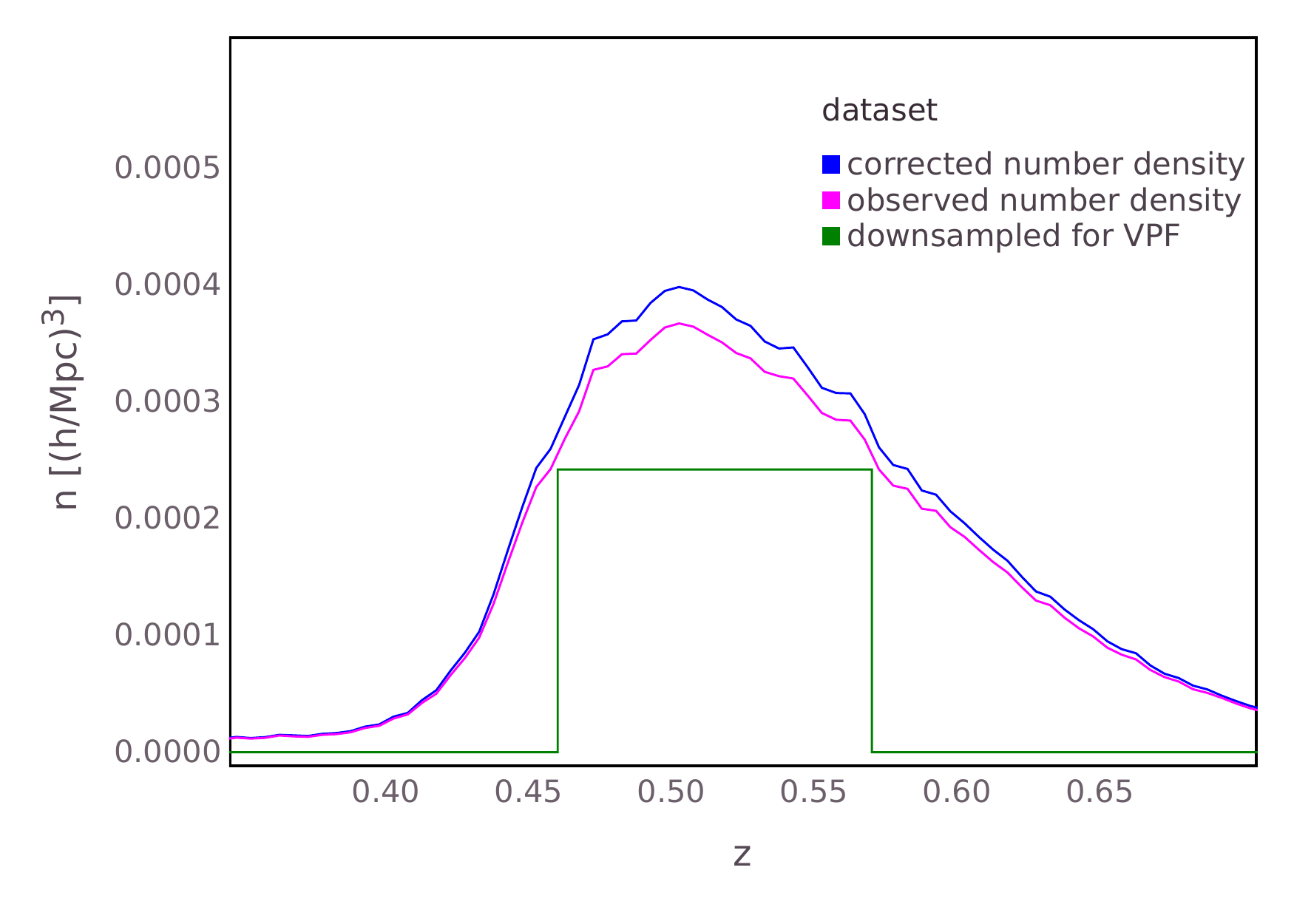}
    \caption{
       This figure shows the number densities as a function of redshift for the observed
        CMASS NGC galaxy sample, the corrected number density when accounting for fibre collisions,
        and the sample used for the VPF measurement that has a constant number density within the
        redshift range $z = [0.4575, 0.5725]$.}
    \label{fig:n_z}
    \end{center}   
\end{figure}

Our probe of void statistics is the void probability function (VPF)
which we will define presently. The VPF is highly sensitive to sample
number density, thus we restrict the BOSS data to a range of redshift
for which the number density is above some threshold. We then randomly
subsampled the data within this redshift interval such that the
remaining galaxies would have a constant number density at comoving
volume in redshift bins in that range. The redshift range we chose for
the Northern Galactic Cap (NGC) CMASS (the high-redshift portion of
BOSS targets) galaxies was $z = [0.4575, 0.5725]$ such that the number
densities of the remaining galaxies was 65\% of the peak number
density of the sample. These values were chosen as a compromise
between maintaining a high number density while also probing a large
volume of the sample by not overly restricting the redshift range. The
resulting galaxy sample has just under 500,000 galaxies. We randomly
subsample the galaxies at each location in redshift, rather than a
more complicated selection, due to that fact that the clustering of
CMASS galaxies is roughly constant with redshift
(\cite{white_etal:11}). Figure~\ref{fig:n_z} shows the number density
of the downsampled galaxies as a function of redshift, compared with
the original sample and with the corrected number density, which is
discussed in the following section.

\subsection{Correlation Function Measurements}
\label{sec:wp}

To quantify two-point clustering we use the projected correlation
function $w_p(r_p)$ \citep{DavisPeebles}. This function is defined as
the integral of the two-dimensional clustering measurement
$\xi(r_p, \pi)$ along the line-of-sight direction $\pi$ and is
insensitive to redshift space distortions, while offering good
constraints on HOD parameters through its signals on small projected
scales $r_p$:

\begin{equation}
    w_p(r_p) = 2 \int_0^{\pi_{\rm max}} \xi(r_p, \pi) d\pi, 
\end{equation}

\noindent with $\pi_{\rm max}=80$ $\,h^{-1}{\rm Mpc}$. We use the Landy-Szalay estimator
\citep{LandySzalay} for two-point statistics. We also choose 24 logarithmic spaced bins in $r_p$ in the 
range $[0.3, 60.0] \,h^{-1}{\rm Mpc}$.

Due to the size of spectroscopic fibres used to take the galaxy redshift data,
and the limited number of overlapping regions in the survey geometry, a
significant fraction of target BOSS galaxies were subject to fibre collisions
where only one galaxy in a collided pair was observed (more details in e.g. 
\citealt{AndersonDR11}, \citealt{Ross17}).
In our measurements, it was necessary to correct for the missing galaxy pairs
that lie within the fibre collision angular radius of $62''$, which corresponds
to a projected separation of around $r_p \sim 0.4 \,h^{-1}{\rm Mpc}$ at the CMASS mean
redshift and thus has a big impact on small scale clustering. To estimate the
true clustering, we use the method of \citep{white_etal:11} and 
upweight the pair counts of any pair found within that separation to account for 
the missing pairs. The extra weighting factor was determined by finding the
ratio of angular clustering measurements $w(\theta)$ between the BOSS galaxies
and the photometric SDSS data from the same field. We find we should use a
factor of 2.63 to upweight pairs whose galaxies are within $62''$ of one
another. The corrected
number density of the CMASS galaxies, after accounting for the missing galaxies
due to fibre collisions and other systematic factors, is shown in
Figure~\ref{fig:n_z}.

To make a direct comparison of the galaxies and underlying dark matter 
distribution with both the $w_p$ and VPF measurements, we restricted the
galaxies to the same redshift range as is used for the VPF measurement before
counting the pairs. We don't randomly subsample the galaxies as is done for the
VPF sample. A random sample of galaxies will still have the same clustering
properties as the full original sample and therefore this extra step would only
increase uncertainty in the $w_p$ measurement. The results of this measurement
are shown in Figure~\ref{fig:wp_vpf}, along with the measurements of the VPF,
described in the next section.

\subsection{VPF Measurements}
\label{sec:vpf}

The void probability function measurement  $P_0(r)$ is made by randomly placing
spheres of some radius $r$ in many locations of the survey and counting the
fraction of empty spheres. Assuming shot noise error in the number of empty 
spheres found, an appropriate number of spheres must be placed to make a
reasonable determination of $P_0$. As sphere scale $r$ increases, empty spheres
become increasingly rare and more spheres are required to find any signal. Based
on the rarity of void spheres above $35 h^{-1}Mp$ within our survey volume, and
the computational cost of searching sufficient spheres at that scale, we have
made a VPF measurement for voids in the range $[5, 35] \,h^{-1}{\rm Mpc}$ in steps of
$5 \,h^{-1}{\rm Mpc}$, using random samples of up to $10^7$ spheres at each radius. 

\begin{figure}
    \begin{center}
    \includegraphics[width=\columnwidth]{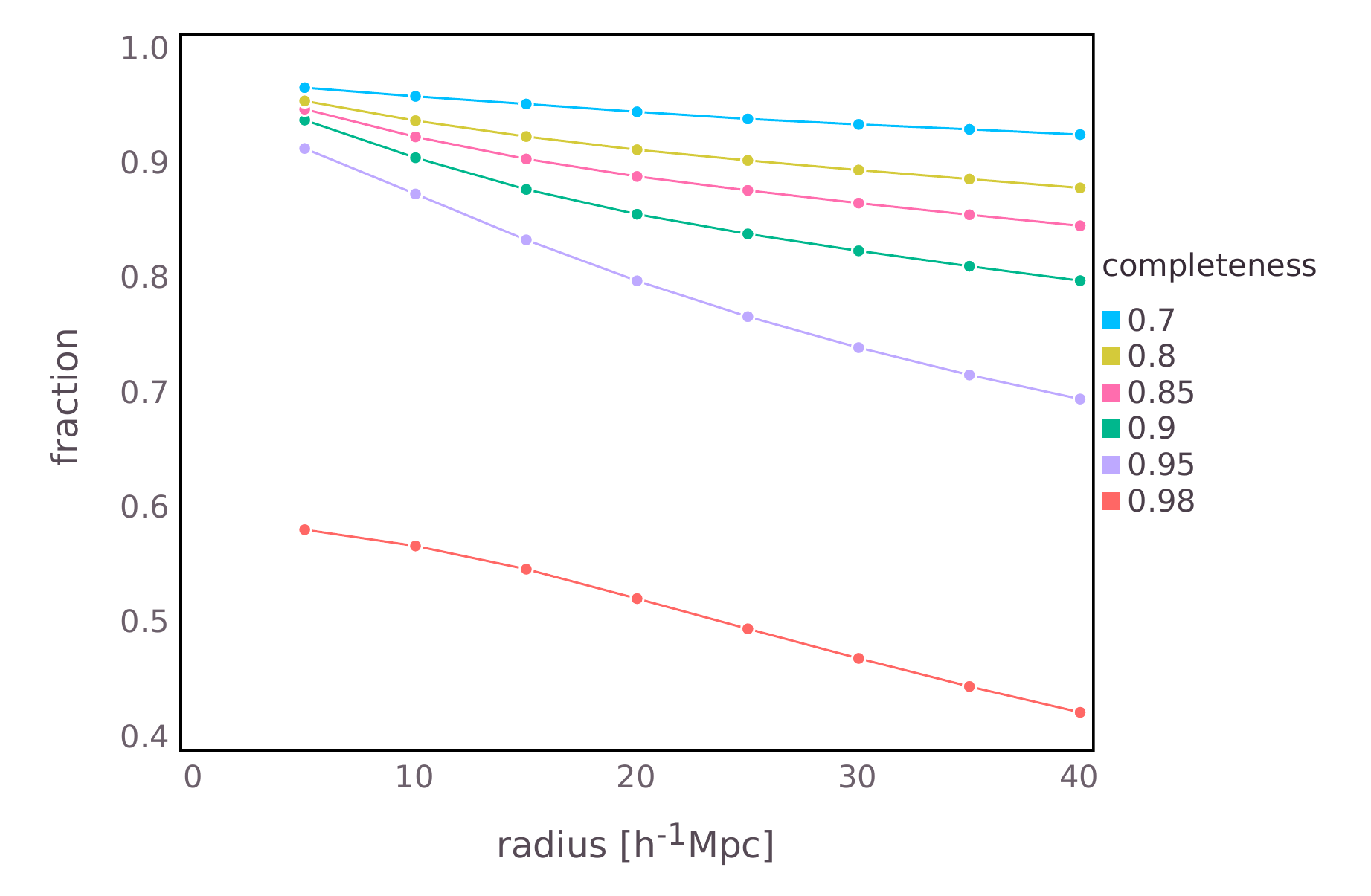}
    \caption{
        As a test for an appropriate completeness threshold, we test the fraction
        of spheres which meet a given threshold in the CMASS survey geometry.
        This plot shows the fraction as a function of sphere radius for a range of thresholds, as described in Section~\ref{sec:vpf}.
        }
    \label{fig:completeness}
    \end{center}
\end{figure}

Due to the nature of the survey geometry, which has irregular borders
along with missing points in its interior due to stars and other
features, not all possible spheres with centres lying within the
survey mask will be fully contained within the survey. As such, care
must be taken to ensure that a given sphere can be said to truly be a
void sphere. Thus, before assessing whether spheres are devoid of
galaxies or not, we must first assess whether they are valid survey
spheres. In order to be considered as such, we require any sphere that
is accounted for in the VPF measurement to meet a completeness
threshold, whereby a minimum volume of the sphere is found to be
within the survey. We tested the fraction of conserved spheres at each
radius for a range of completeness thresholds by randomly proposing
spheres within the survey mask and estimating the percentage of the
sphere volume that lay outside the mask by using random "veto points"
placed at the survey edges and in the interior points that were
compromised. The results after averaging the volume over $10^7$
spheres at each radius are shown in Figure~\ref{fig:completeness}. We
decide that spheres up to 95\% completeness result in a sufficiently
large fraction being conserved for the VPF measurement, where over two
thirds of spheres are still valid even up to $40$ $h^{-1}{\rm Mpc}$. We will
describe our method of making proper comparisons between the mocks and
data in the next section.

\begin{figure*}
    \begin{center}
    \begin{minipage}{\textwidth}
    \includegraphics[width=\textwidth]{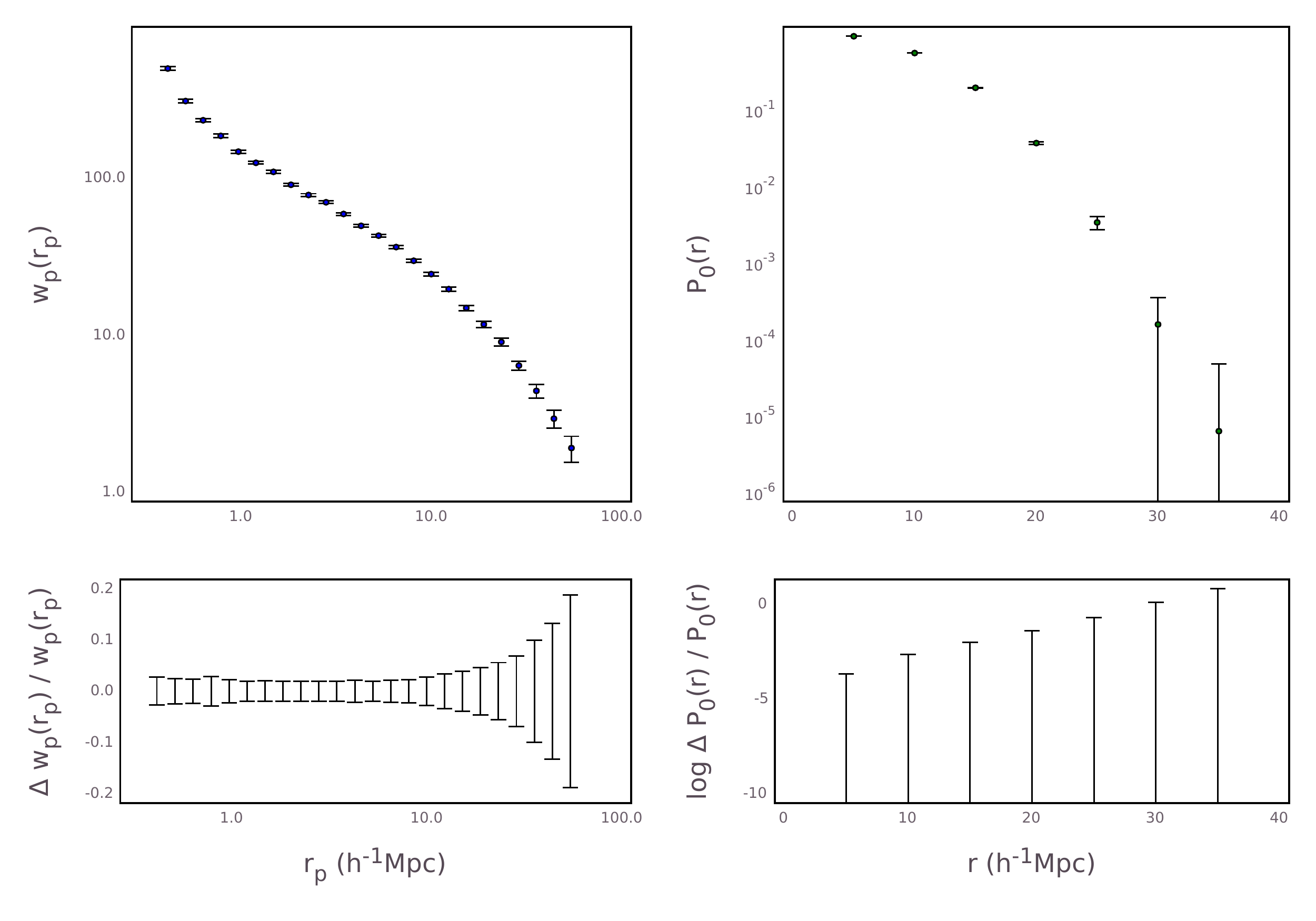}
    \caption{
        This figure shows the projected correlation function $w_p$ with fractional
        errors and void probability function with fractional errors. The errors of
        the VPF are shown in log space since they increase effectively 
        exponentially with void radius.}
    \label{fig:wp_vpf}
    \end{minipage}
    \end{center}   
\end{figure*}

The results for the CMASS northern sample are shown in
Figure~\ref{fig:wp_vpf} along with the $w_p$ measurement for the same
galaxies. Errors are described in the following subsection.

\subsection{Estimating Covariances with Mock Samples}

We determine an estimate of the covariance in our measurements by
using a set of mock galaxy samples that were made for the purpose of
constructing covariance matrices for the CMASS survey data. The mocks
used were the quick particle mesh (QPM) mocks \citep{QPM} created for
BOSS. We made the measurements to match the methods described in the
previous sections exactly, using the same estimators and bins and
radii, and found the full covariance of these measurements over a set
of 400 mocks. The covariance matrix we calculate using these mocks is
used for making parameter inference of the HOD model, as described
below. The diagonal variances of the matrix were used to determine the
errorbars in the measurements shown in Figure~\ref{fig:wp_vpf}. For
the VPF, we show the fractional error bars in log-space, as the range
is very broad. At small scales, the VPF does not contain significant
information about large scale structure and the errors are consistent
with simple Poisson errors. However, at scales larger than the mean
galaxy spacing (roughly 16 $h^{-1}{\rm Mpc}$) the amplitude of the VPF
is reflective of the structure, but the fractional error bar is
significantly larger. This is both due to the rarity of empty spheres
at these scales and sample variance in the clustering of galaxies.

\section{Galaxy Clustering and Voids from a Halo Occupation Model}
\label{sec:vpf_mod}

\subsection{Dark Matter Simulation and Halo Catalogue}

To compare the measurements with models of halo occupation, we use a
halo catalogue from a high resolution dark matter simulation. For this
work use the Big MultiDark Planck (BigMDPL) simulation \citep{MDPL,
  MultiDark}. The simulation has $2.5 h^{-1} Gpc$ comoving volume and
$3840^3$ dark matter particles, giving it a mass resolution of
$2.359 \times 10^{10} h^{-1} M_{\odot}$. It uses a Planck cosmology
\citep{Planck} with $h = 0.6777$, $\Omega_{\Lambda} = 0.693$,
$\Omega_M = 0.307$, and $n = 0.96$.  The simulation was performed
using GADGET-2 code \citep{GADGET} and further details can be found in
\citet{Klypin2016}. The Planck cosmological parameters are used
throughout this work to remain consistent with the simulation.

The MultiDark Database also provides halo catalogues for each
simulation, which we use for our analysis. The catalogues were made
using the Rockstar Halo Finder \citep{Rockstar}. We use the host halos
from this catalogue -- those not contained within the virial radius of
any other halo as subhalos -- for the purposes of our model. The halos
have masses $M_{200}$ determined from their spherical overdensity,
measured for spheres of $\Delta = 200$ times the critical density. To
compute the halo concentration for the halos in the catalogue, we use
the empirical relationship found in \citet{DuttonMaccio} for halo
masses of this definition at a redshift of $z = 0.5$.

We have also received dark matter particle data from the MultiDark
authors with which to measure the dark matter overdensity at
$15 \,h^{-1}{\rm Mpc}$ scales at the locations of each of the host
halos. We use these halo properties, along with their relative
positions and velocities in the simulation, to build model galaxy
catalogues with which to estimate galaxy clustering.

\subsection{Halo Occupation Distribution}

To construct model galaxy catalogues from the halos, we use the Halo
Occupation Distribution (HOD) described in~\citet{Zheng07}. This is a
statistical prescription for the number of central and satellite
galaxies that live in a given halo, depending on the mass of that
halo. For centrals, there can be either zero or one galaxy, with the
mean central occupation at any mass given by

\begin{equation}
    \label{eq:cen_occ}
    \langle N_{\rm cen} \rangle_M = \frac{1}{2} \left[ 1 + \erf \left( \frac{\log M - \log M_{\rm min}}{\sigma_{\log M}} \right) \right],
\end{equation}

\noindent where $M_{\rm min}$ is the halo mass at which the
probability for a galaxy existing at its centre goes from less than a
half to greater than a half, and $\sigma_{\log M}$ effectively
determines the scatter in halo mass between halos with no central
galaxies and halos which all have a central galaxy. If a halo has a
central galaxy, drawn from a Bernoulli distribution with probability
$\langle N_{\rm cen} \rangle$ the galaxy is given the position and
velocity of the halo centre of mass.

The average number of satellite galaxies in a host halo of mass $M$ is
given by

\begin{equation}
    \label{eq:sat_occ}
    \langle N_{\rm sat} \rangle_M = \frac{1}{2} \left[ 1 + \erf \left( \frac{\log M - \log M_{\rm min}}{\sigma_{\log M}} \right) \right] \left[ \left( \frac{M}{M_1} \right)^{\alpha} e^{- \frac{M_{\rm cut}}{M}} \right],
\end{equation}

\noindent which is a slight modification of the original Zheng
formulation. It consists of a power law in halo mass, scaled by $M_1$
with slope $\alpha$, modulated by the central occupation relationship
and with an extra exponential cutoff $M_{\rm cut}$. The satellite
number for each individual halo is drawn from a poisson distribution
with mean $\langle N_{sat} \rangle$ (only if the halo has a central
galaxy) and these galaxies are placed around the halo centre according
to an NFW profile with the halo's concentration. The satellites are
given velocities assuming an isotropic distribution of virialized
objects in the halo's gravitational potential.

\subsection{Modeling Assembly Bias}
\label{sec:vpf_assembias}

To determine whether there is any environment-dependent bias in the
galaxy occupation of dark matter halos which is independent of the
halo mass, we use the dark matter overdensity of host halos. We
measure the relative density $\rho$ of a halo by counting the number
of dark matter particles in a sphere of $15 \,h^{-1}{\rm Mpc}$ located
at the halo's centre of mass and dividing by the average number
expected over the whole simulation, for a random subsample of dark
matter particles for the simulation. \cite{tinker_etal:06_voids} and
\cite{tinker_etal:08_voids} used a simple model where halo occupation
changed by a set amount at densities below some critical density. The
results of \cite{tinker_etal:06_voids} demonstrated that a model of
this kind could produce changes to both $w_p(r_p)$ and the VPF, but that
the VPF was far more sensitive to these changes and could be detected
to high significance even when the change to $w_p(r_p)$ was negligible
within the statistical precision of the measurements.

Here, we extend the method of \cite{tinker_etal:06_voids} to increase
the flexibility of the model. We modify the HOD described above my
changing the value of $\log M_{\rm min}$ as a function of $\rho$. For
each halo, we add an offset to $\log M_{\rm min}$ of

\begin{equation}
  \label{e.assbias}
    \Delta_{M_{\rm min}} = \frac{f_{\rho}}{2} \left[ 1 + \erf \left( \frac{\log \rho - \log \rho_{\rm th}}{\sigma_{\rho}} \right) \right].
\end{equation}

\noindent This leads to a change in the minimum halo mass for which a halo will
be occupied by central galaxies and also changes the modulation of the
satellite galaxy occupation by the same factor. The value for
$M_{\rm min}$ is effectively changed to have two regimes of halo
occupation -- one value at higher densities, and another value at
lower densities, with a difference in log of $f_{\rho}$ that is
separated by a threshold density $\rho_{\rm th}$ with a soft
transition dictated by $\sigma_{\rho}$. This model allows us to assess
whether the data allow for a mechanism of galaxy-formation that
prefers galaxies to be formed in halos of the same mass when they
occur in different environments.

\begin{figure*}
    \begin{center}
    \begin{minipage}{\textwidth}
    \includegraphics[width=\textwidth]{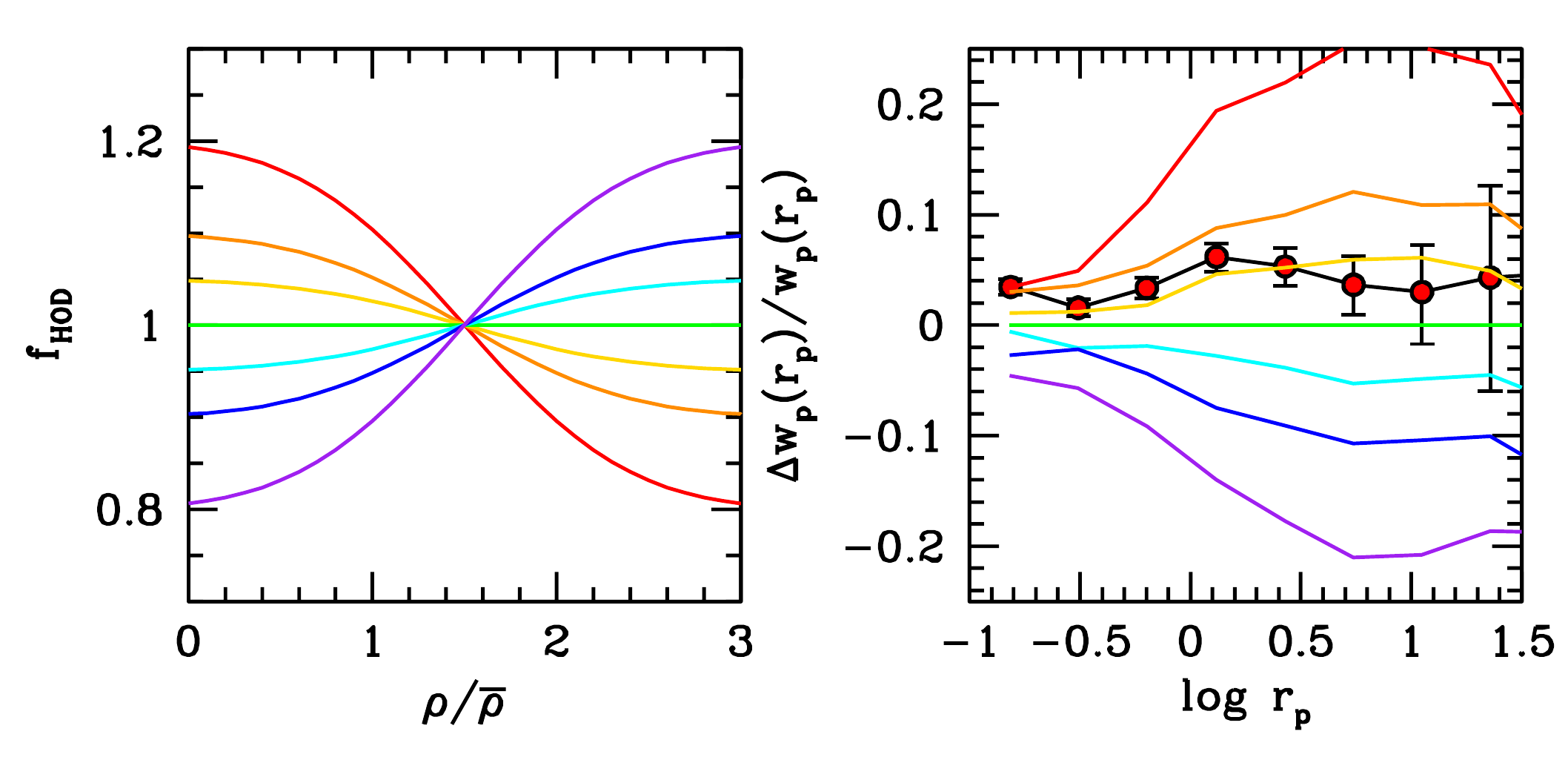}
    \caption{ Our model for incorporating galaxy assembly bias. The
      left-hand panel shows several implementations of equation
      (\ref{e.assbias}). Here $f_{\rm HOD}$ is the fractional change
      to the HOD mass scale. The parameters $\sigma_p$ and
      $\rho_{\rm th}$ are held constant, but $f_\rho$ is varied to
      both increase and decrease galaxy formation efficiency at
      densities below the mean density around CMASS galaxies. The
      right-hand panel shows the impact these variations have on the
      two-point clustering of galaxies. The points with errorbars show
      the amount of assembly bias predicted in a model where galaxy
      mass is matched to peak circular velocity of the halo, which is
      correlated with halo formation history. The points themselves
      show the clustering of this model relative to a shuffled version
      of the same model, where galaxies are shuffled randomly between
      halos of the same mass, thereby erasing any correlations with
      the large-scale density field, but preserving the HOD. The
      expected assembly bias is small, and can be easily modeled in
      our framework.}
    \label{assbias}
    \end{minipage}
    \end{center}   
\end{figure*}

The choice of parameterization in Equation (\ref{e.assbias}) is meant
to give maximum flexibility to the model. It allows galaxy formation
efficiency to both increase or decrease in voids, it changes the
density scale at which this change can happen, and the rapidity with
which the change happens along the edge of the void. Figure
\ref{assbias} shows several examples of how shifts in the mass scale
impact clustering. We also compare these results to a model of
assembly bias where galaxies are assigned to halos using the abundance
matching approach, with galaxy luminosity matched to the peak maximum
circular velocity over the history of each halo (see
\citealt{wechsler_tinker:18} for a review of abundance matching
methods). Peak circular velocity is also correlated with large scale
density at fixed halo mass, thus imprints an assembly bias signal on
the galaxy population without explicitly using the large-scale
density. However, the change to the clustering created by this
abundance matching model is easily within the parameter space of
equation (\ref{e.assbias}). But we don't wish the restrict ourselves
to only the types of assembly bias created in this way---thus, the
increased flexibility of equation (\ref{e.assbias}) is more
appropriate for the purposes of this study.

\subsection{Modeling Galaxy Clustering}

\subsubsection{Projected Correlation Function}

Using the halo catalogue and HOD described above, we can simulate a galaxy
catalogue which we can use to make model measurements to compare with the
measurements made in the data. As described in Section~\ref{sec:wp}, the
projected correlation function measurement made in the CMASS sample has been
corrected for any geometric effects and fibre collisions through the use of a
carefully constructed random point distribution over the survey mask and with
collision pair-upweighting. Therefore, we need not make any special
considerations in our model measurement. We make a distant observer
approximation for the simulation box and take one dimension
to be the line-of-sight axis. Along this dimension $x_3$, we displace each 
galaxy according to its parallel velocity $v_3$ by

\begin{equation}
    \Delta x_3 = \frac{1 + z}{H(z)} v_3
\end{equation}

to mimic redshift-space distortions, where $H(z)$ is the Hubble parameter at the
simulation redshift $z$.

We use the updated galaxy positions to calculate the pairs for the $\xi (r_p, \pi)$ 
estimator with $\pi$ along the $x_3$ dimension of the box. Since the box is 
periodic, we can analytically compute the expected random pairs for the estimator
by assuming a poisson process to save on computational time. The code we use for
calculating the theory $w_p(r_p)$ is \texttt{Corrfunc} \citep{Corrfunc}.

\subsubsection{Void Probability Function}

In order to make an honest comparison of the simulated galaxies to the observed
galaxies with respect to the VPF, we must first match the galaxy number density.
We downsample the simulated galaxies randomly until we reach the same constant
number density as we had for the redshift range for the VPF measurement
discussed in Section~\ref{sec:dataprep} and shown in Figure~\ref{fig:n_z}.

Since the measurement of the VPF made in the data was made using a sample of 
galaxies with missing observations due to fibre collisions, it was necessary to 
also simulate such fibre collisions in our mock galaxies. Otherwise,
downsampling the simulated galaxy sample to our chosen number density would lead
to more voids, since there would be no preferential removal of collided pairs,
which are typically in denser regions. Removing "fibre collisions" first in our
simulated mock leads to a smaller fraction of voids at increasingly larger
radii.

The comoving distance from the nearest galaxies to the furthest in our observed 
sample is approximately $300$ $h^{-1}{\rm Mpc}$. The depth of the simulation box, along
our line of sight dimension, is far greater, at $2500$ $h^{-1}{\rm Mpc}$. The number of
galaxy pairs that will seem to be "collided" by the fibre collision radius of
our simulation redshift (which comes to $\sim 0.4$ $h^{-1}{\rm Mpc}$ in comoving
separation) is thus also far greater in the box than in the survey. It becomes 
an issue to obtain the desired number density in our mock because there can be 
enough collided galaxies, that removing the same fraction of them as are
unobserved in the data, leads to a lower
mock number density than the final data sample. As a workaround, we restrict the
simulation to a fraction of its size along the line-of-sight dimension. To be as
consistent with the survey data as possible, we choose the fraction such that 
the ratio of simulation box "depth" to "area", is similar to the ratio of survey
depth and area. The full survey volume is approximately $10^9 h^{-3} Mpc^3$, so
we choose
our effective depth for the simulation box to be $600$ $h^{-1}{\rm Mpc}$ and make the
measurement in this restricted portion of the mock volume.

Once we have removed collided mock galaxies appropriately and sampled the mock 
down to match the observed constant number density sample, we measure the VPF in
the mock. We make the measurement with $10^7$ spheres at the largest radii, as
in the data, and use random points distributed evenly throughout the mock. The
mock is assumed periodic in the transverse dimensions, but voids are kept one
sphere radius from the edges in the restricted line-of-sight dimension.

\subsection{Fitting HOD Parameters to the Measurements}

To fit the model values for $w_p$ and $P_0(r)$ to the observed values,
we use a standard $\chi^2$ statistic. We compute the statistic for
samples in the HOD parameter space of $M_{\rm min}$,
$\sigma_{\log M}$, $M_1$, $M_{\rm cut}$, and $\alpha$ (and including
$f_{\rho}$, $\rho_{\rm th}$, and $\sigma_{\rho}$ for the density
dependent model). For each sample, we populate the halo catalogue with
galaxies accordingly and then compute the simulated measurements. The
samples are drawn from a posterior distribution obtained by
multiplying the $\chi^2$ values with prior probabilities, chosen to
have flat distributions within physically reasonable
ranges.

Any implementation of an HOD requires the number density of the sample
to be modeled. For the CMASS sample, which has a variable number
density, the choice of which density to use is not entirely
clear. Here we use the mean corrected number density within the
redshift range of our sample. The sample variance on this value is
very small, at the percent level, but we allow the number density of
the HOD model to vary by $\pm 10\%$ relative to the mean. This
incorporates the possibility that the mean value is not the most
appropriate, and marginalizes over other possibilities. For each mock,
however, the VPF is always measured after downsampling the mock to the
same value as the CMASS sample on which we make our VPF measurements.
The posteriors are sampled using a Markov Chain Monte Carlo (MCMC)
sampler based on an affine-invariant search algorithm
\citep{GoodmanWeare}.

We perform the fit for multiple configurations of the model and
measurements.  In the first experiment, we fit only to the $w_p$
measurement, using the standard, non-density dependent HOD, to get a
fiducial set of posteriors and compare the VPF distribution of the
mocks from the posterior samples to the observed VPF. We make another
fit with the standard HOD, fitting to both $w_p$ and the VPF
$P_0$. Finally we make a fit using the density-dependent HOD, fitting
to both the $w_p$ and $P_0$.

\section{Results}
\label{sec:vpf_res}

\begin{figure*}
    \begin{center}
    \begin{minipage}{\textwidth}  
    \includegraphics[width=\textwidth]{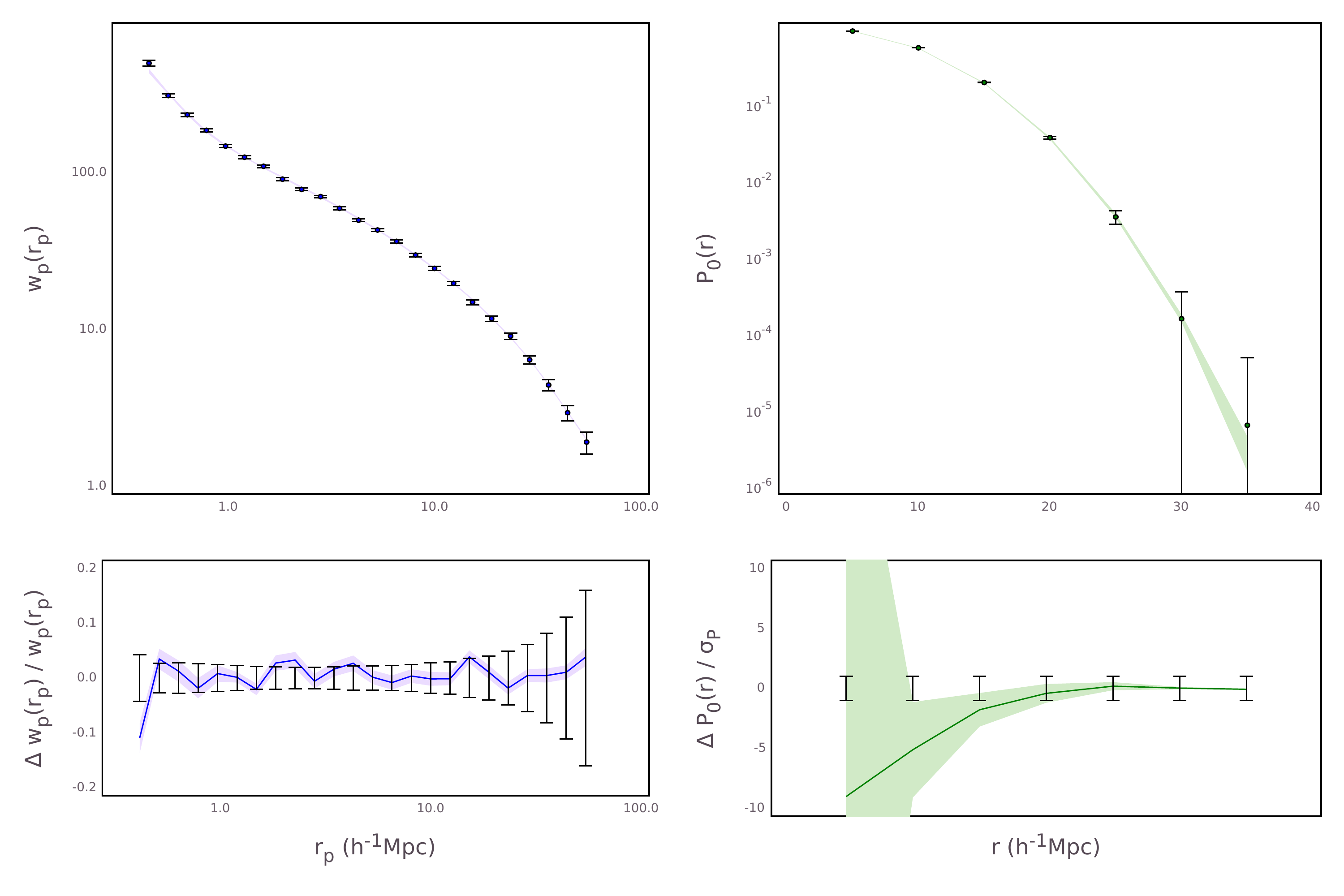}
    \caption{
        This figure shows the same measurements as in Figure~\ref{fig:wp_vpf} and
        includes the predicted values from mock measurements in simulated galaxy
        samples. The mock galaxies are derived from parameter samples from the 
        posterior over the HOD parameters, obtained from fitting the standard HOD
        using the $w_p$ measurement. The model predictions show a region within one
        standard deviation of the predictions from the posterior. In the residual
        plot for the VPF (lower right panel) we compare model predictions as a
        distance in standard deviations (i.e. the observation errors) in order to 
        clearly show the values in each bin.
        }
    \label{fig:wp_vpf_compare}
    \end{minipage}
    \end{center}   
\end{figure*}

The comparison of posterior predictions for both $w_p$ and $P_0$ for
the fiducial fit (constrained using only $w_p$) are shown in
Figure~\ref{fig:wp_vpf_compare}. The fiducial model provides a good
fit to the VPF, despite only using the information of the two-point
function in the data to constrain the model.  This suggests already
that there is no need for the inclusion of density dependence
in the galaxy-halo model to describe galaxy clustering statistics that
are sensitive to a range of densities within the cosmic web.

\begin{figure*}
    \begin{center}
    \begin{minipage}{\textwidth}  
    \includegraphics[width=\textwidth]{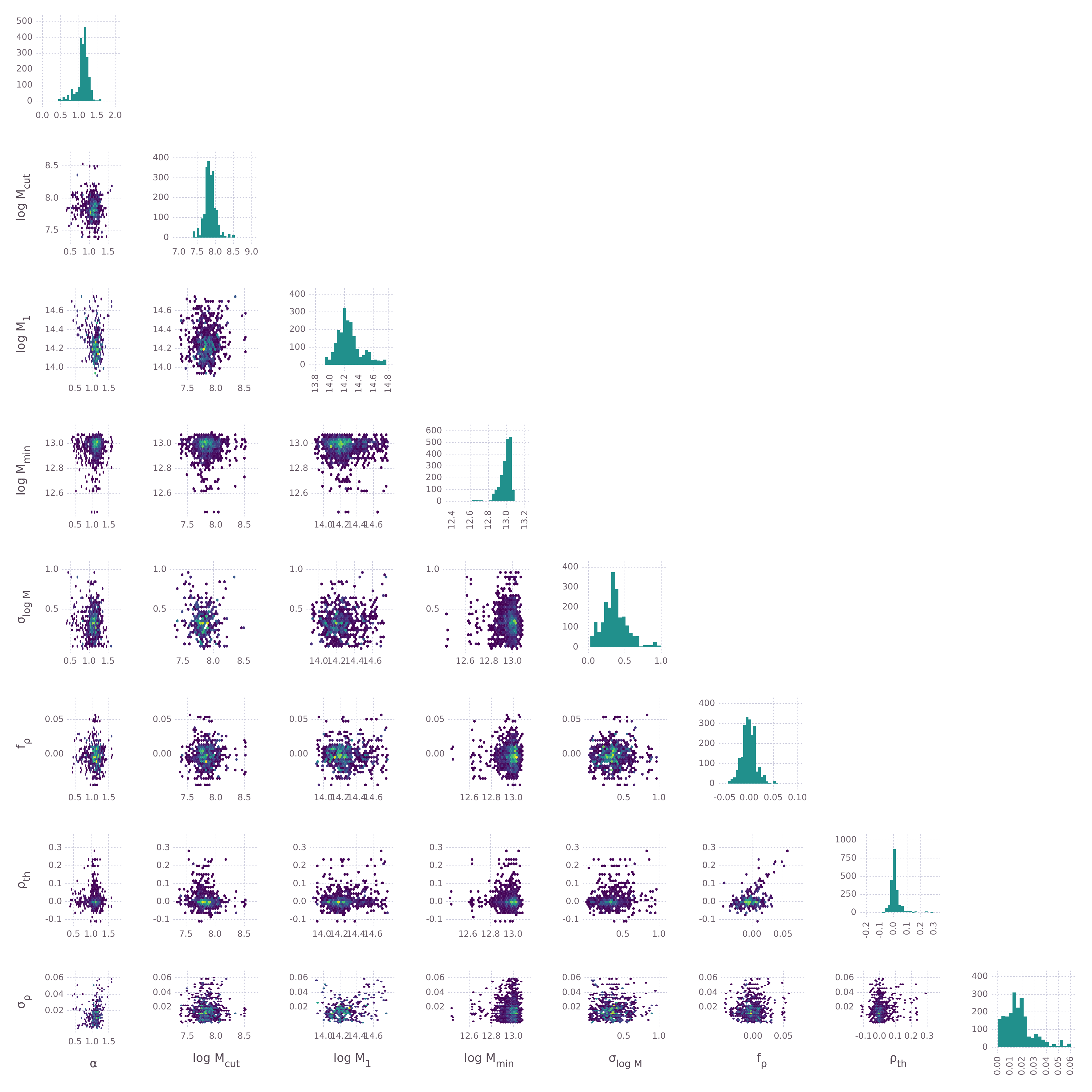}
    \caption{
        Here we show the posterior samples for the model described in 
        Section~\ref{sec:vpf_assembias} when fitting to $w_p$ and $P_0$ from the
        CMASS northern sample including assembly bias parameters. 
        }
    \label{fig:assembias_posterior}
    \end{minipage}
    \end{center}   
\end{figure*}

\begin{figure*}
    \begin{center}
    \begin{minipage}{\textwidth}  
    \includegraphics[width=\textwidth]{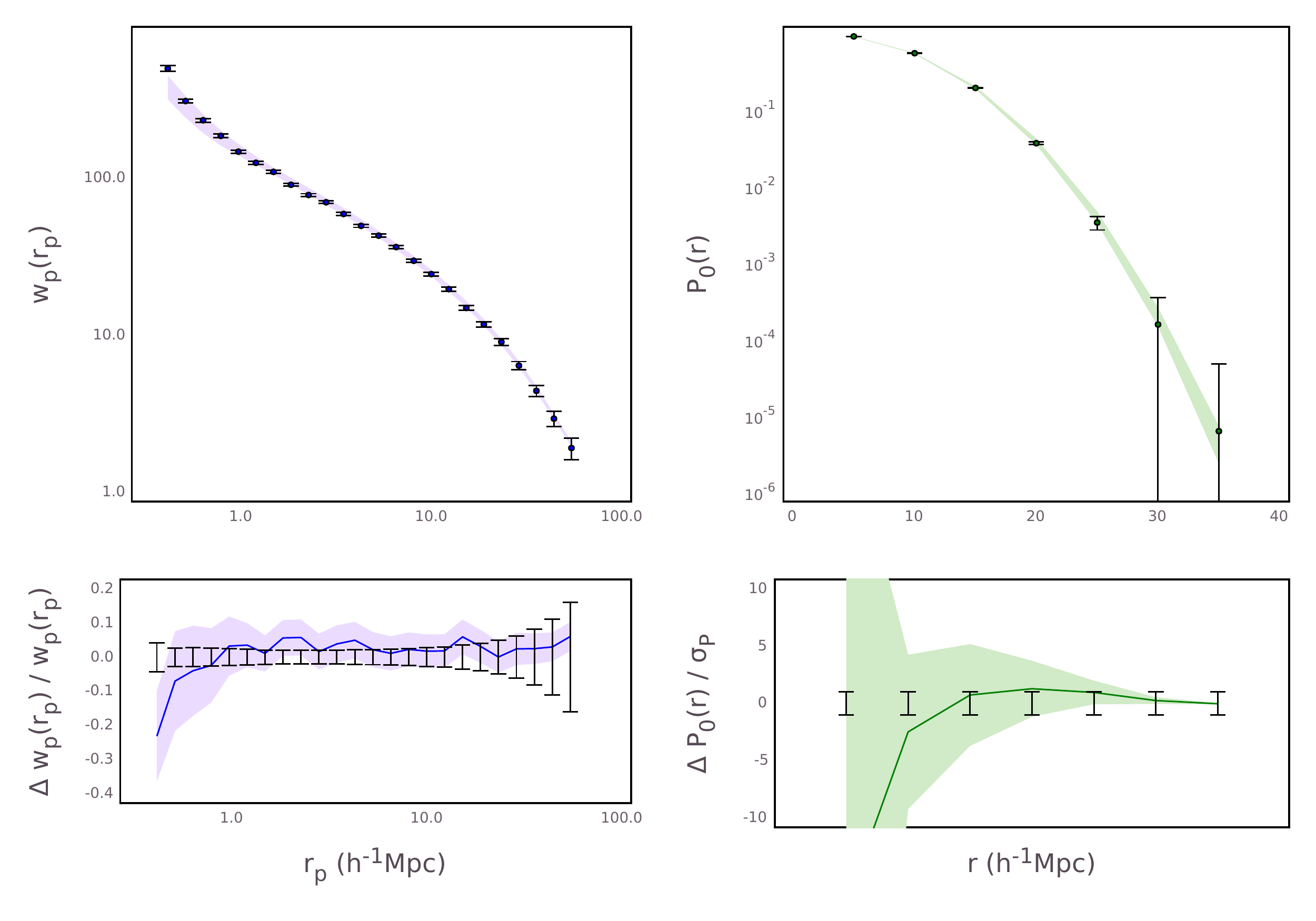}
    \caption{
        This figure is analogous to what is shown in 
        Figure~\ref{fig:wp_vpf_compare},
        however, in this case, the mock results are derived from posterior 
        predicted galaxy samples from a HOD model incorporating density
        dependence, using parameters sampled from the posteriors shown in 
        Figure~\ref{fig:assembias_posterior}
        }
    \label{fig:wp_vpf_delta}
    \end{minipage}
    \end{center}   
\end{figure*}

As a confirmation of this lack of density dependence in halo
occupation, we show the posterior parameter distributions of a model
including density dependence -- as described in
Section~\ref{sec:vpf_mod} -- in
Figure~\ref{fig:assembias_posterior}. The density parameters,
particularly the offset magnitude $f_{\rho}$ are all consistent with
having no impact on the model. An analogous comparison of the
predictions for $w_p$ and $P_0$ as in Figure~\ref{fig:wp_vpf_compare},
but made using the HOD including this density dependence, is shown in
Figure~\ref{fig:wp_vpf_delta}. Unsurprisingly, there is similarly good
correspondence with the data.

A table summarizing the maximum a posteriori values along with the
standard deviation of marginalized samples in each parameter, for each
of the model cases, is shown in Table~\ref{tab:vpf_constraints}. This
includes the test where we simultaneously fit both the $w_p$ and $P_0$
for the density-independent model. As expected from Figure
\ref{fig:wp_vpf_compare}, the parameter constraints are consistent
with the model in which only $w_p$ is fit. The only significant
difference is in $M_{\rm cut}$, but the difference is only in the
numerical values---at such small values of $M_{\rm cut}$, the cutoff
in the satellite occupation function has no quantitative impact on the
clustering because it is so much smaller than $M_{\rm min}$. The
agreement on the HOD parameters extends to the density-dependent model
as well. The best-fit value of $f_\rho$ is $0.02\pm 0.04$, clearly
consistent with zero---i.e., no density dependence. Indeed, values
significantly away from zero are strongly excluded by the data.

\begin{table*}
    \newcolumntype{d}{D{,}{,}{-1}}
	\centering
	\caption{
	    This table shows the top hat prior ranges (in some cases improper) along with the maximum
	    a posteriori values and standard deviations from the posterior
	    chains of our models. The three sets of posteriors are taken from a model fit only to the
	    two-point function, a model with mass-only HOD fit to both the two-point function and VPF,
	    and a model fit to both and incorporating density dependence in the HOD. The constraints
	    are consistent across all models, showing that a mass-only model is sufficient to fit the
	    data and that assembly bias in the BOSS LRGs is strongly ruled out.}
	\label{tab:vpf_constraints}
	\begin{tabular}{cdS[table-format = -2.2(2)]S[table-format = 2.2(2)]S[table-format = -2.2(2)]}
		\hline
        {\rm parameter} & {\rm priors} & \ensuremath{w_p} only & \ensuremath{w_p} \& \ensuremath{P_0} & \ensuremath{w_p} \& \ensuremath{P_0} with \ensuremath{f_{\rho}}\\
		\hline
		\ensuremath{\alpha}          & [0, 3]                                                           & 1.17(06)     & 1.12(16)  & 1.00(23) \\
		\ensuremath{M_{\rm cut}}     & (\ensuremath{-\infty}, \ensuremath{\infty})                      & -17.47(1150) & 4.87(17) & 8.00(29) \\
		\ensuremath{M_1}             & [\ensuremath{M_{\rm min}}, 15]                                   & 14.2(03)     & 14.28(17) & 14.22(17) \\
		\ensuremath{M_{\rm min}}     & [10.5, 15]                                                       & 13.07(04)    & 13.18(13) & 12.92(14) \\
		\ensuremath{\sigma_{\log M}} & (0, 1.5]                                                         & 0.35(06)     & 0.55(13)  & 0.45(25) \\
		\ensuremath{f_{\rho}}        & [\ensuremath{10.5 - M_{\rm min}}, \ensuremath{15 - M_{\rm min}}] &              &           & 0.02(04) \\
		\ensuremath{\rho_{\rm th}}   & [0.5, 1.5]                                                      &              &           & 1.02(11) \\
		\ensuremath{\sigma_{\rho}}   & (0, \ensuremath{\infty})                                         &              &           & 0.03(01) \\
		\hline
	\end{tabular}
\end{table*}

\section{Discussion}

We have shown that the information on galaxy bias encoded in the
galaxy two-point correlation function is sufficient to model galaxy
bias in extremely low-density environments. Thus, halo occupation does
not change from high to low densities. Additionally, incorporating a
flexible model for galaxy assembly bias---a model in which halo
occupation depends on both $M_h$ and large-scale density---only fits
the combined $w_p$ and VPF data when the density-dependent parameters
are consistent with zero. Studies meant to extract cosmological
information from non-linear clustering of large-scale galaxies
redshift surveys, like BOSS, it's successor eBOSS (\cite{dawson_etal:15}) and
the near-term DESI survey (\cite{desi_fdr}), can expect minimal-to-no
degeneracies between cosmology and assembly bias.

This result appears to be in tension with those of \citet{Saito16},
where measurements of $w_p$ are better fit using an age-matching model
for galaxy occupation in a halo-abundance matching context. However,
the results of \citet{Saito16} are mostly inferred from
redshift-dependent effects in their model, which we do not encounter,
given the relatively narrow redshift range of this work. Another
difference is our implementations of assembly bias, which in our model
is done explicitly through a dependence on $\rho$, and in theirs done
implicitly through abundance matching. However, Figure \ref{assbias}
shows that the type of assembly bias imparted in this model is within
our parameter space. Additionally, \cite{goh_etal:19} find that the
properties of dark matter halos, once controlling for large-scale
density, are independent of the details of the halo's location within
the cosmic web---i.e., filaments, walls, nodes, sheets, etc. Thus
there is not more information beyond $\rho$ required when implementing
assembly bias.

The ability of the density-{\rm in}dependent HOD model to match {\rm
  both} the clustering and void distribution is somewhat surprising
given recent results showing that a mass-only approach has difficulty
yielding statistically acceptable fits to the galaxy correlation
function for low-luminosity samples in SDSS (\citealt{reddick_etal:13,
  zentner_etal:16, lehmann_etal:17}). All of these works found a
better fit when correlating halo occupation with something other than
mass, either through an explicit second parameter (i.e., the
`decorated HODs' of \citealt{hearin_etal:16_decorated}), or by
abundance matching on galaxy properties other than mass. However, both
\cite{zentner_etal:16} and \cite{lehmann_etal:17} show no detection of
assembly bias for the brightest samples that they examine: $M_r-5\log
h<-21$ and $-22$, respectively. The CMASS sample is not immediately
comparable to a volume-limited, complete SDSS sample, but the number
density of the CMASS sample is roughly consistent with the number
density of the $M_r-5\log h<-21.5$ sample, given some grounding of the
halo mass scales involved in each sample.

Additionally, the CMASS sample, as well as the brighter SDSS samples,
are comprised almost exclusively of red-and-dead, passive
galaxies. Many studies of color-dependent clustering have shown that
the quenching process cannot correlate with halo formation history
(\citealt{tinker_etal:08_voids, tinker_etal:17_p1, tinker_etal:18_p2,
  zu_mandelbaum:18, sin_etal:17}). Additionally, the small observed
scatter between halo mass and stellar mass for massive galaxies is
most easily explained by a quenching process that depends exclusively
on the galaxy stellar mass (\cite{tinker:17}). Thus, galaxy quenching
may erase any correlation between galaxy mass and halo formation
history that may exist for star-forming galaxies which gave rise to
the positive detections of galaxy assembly bias for lower-luminosity
SDSS galaxies.

Thus, the luminous red galaxy class of targets may be the best
opportunity to obtain cosmological constraints from non-linear galaxy
clustering.




\bibliographystyle{mnras}
\bibliography{refs} 




%
%


\bsp	
\label{lastpage}
\end{document}